\documentclass[%
 aip,
 amsmath,amssymb,
 reprint,%
]{revtex4-1}

\usepackage{graphicx}% Include figure files
\usepackage{dcolumn}% Align table columns on decimal point
\usepackage{bm}% bold math
\usepackage[colorlinks=true,linkcolor=blue,citecolor=blue,urlcolor=blue]{hyperref}
\usepackage[utf8]{inputenc}
\usepackage[T1]{fontenc}
\usepackage{mathptmx}
\usepackage{etoolbox}
\makeatletter
\def\@email#1#2{%
 \endgroup
 \patchcmd{\titleblock@produce}
  {\frontmatter@RRAPformat}
  {\frontmatter@RRAPformat{\produce@RRAP{*#1\href{mailto:#2}{#2}}}\frontmatter@RRAPformat}
  {}{}
}%
\makeatother
\begin{document}

\preprint{AIP/123-QED}

\title{A Voigt laser operating on $^{87}$Rb 780 nm transition}
% Force line breaks with \\
\author{Zijie Liu}
\homepage{These authors contributed to the work eqully and should be regarded as co-first authors}
\affiliation{ 
State Key Laboratory of Advanced Optical Communication Systems and Networks, Institute of Quantum Electronics, School of Electronics, Peking University, Beijing 100871, China%\\This line break forced with \textbackslash\textbackslash
}%
\affiliation{ 
State Key Laboratory of Advanced Optical Communication Systems and Networks, School of Electronics, Peking University, Beijing 100871, China%\\This line break forced with \textbackslash\textbackslash
}%

\author{Xiaolei Guan}%
\homepage{These authors contributed to the work eqully and should be regarded as co-first authors}
\affiliation{%
State Key Laboratory of Information Photonics and Optical Communications, Beijing University of Posts and Telecommunications, Beijing 100876, China%\\This line break forced% with \\
}%

\author{Xiaomin Qin}
\affiliation{ 
State Key Laboratory of Advanced Optical Communication Systems and Networks, Institute of Quantum Electronics, School of Electronics, Peking University, Beijing 100871, China%\\This line break forced with \textbackslash\textbackslash
}%

\author{Zhiyang Wang}
\affiliation{ 
State Key Laboratory of Advanced Optical Communication Systems and Networks, Institute of Quantum Electronics, School of Electronics, Peking University, Beijing 100871, China%\\This line break forced with \textbackslash\textbackslash
}%

\author{Hangbo Shi}
\affiliation{ 
State Key Laboratory of Advanced Optical Communication Systems and Networks, Institute of Quantum Electronics, School of Electronics, Peking University, Beijing 100871, China%\\This line break forced with \textbackslash\textbackslash
}%

\author{Jia Zhang}
\affiliation{ 
State Key Laboratory of Advanced Optical Communication Systems and Networks, Institute of Quantum Electronics, School of Electronics, Peking University, Beijing 100871, China%\\This line break forced with \textbackslash\textbackslash
}%

\author{Jianxiang Miao}
\affiliation{ 
State Key Laboratory of Advanced Optical Communication Systems and Networks, Institute of Quantum Electronics, School of Electronics, Peking University, Beijing 100871, China%\\This line break forced with \textbackslash\textbackslash
}%

\author{Tiantian Shi}
\email{tts@pku.edu.cn}
\homepage{Authors to whom correspondence should be addressed:tts@pku.edu.cn and ahdang@pku.edu.cn}
\affiliation{ 
State Key Laboratory of Advanced Optical Communication Systems and Networks, Institute of Quantum Electronics, School of Electronics, Peking University, Beijing 100871, China%\\This line break forced with \textbackslash\textbackslash
}%

\author{Anhong Dang}
 \email{ahdang@pku.edu.cn}
\homepage{Authors to whom correspondence should be addressed:tts@pku.edu.cn and ahdang@pku.edu.cn}
\affiliation{ 
State Key Laboratory of Advanced Optical Communication Systems and Networks, School of Electronics, Peking University, Beijing 100871, China%\\This line break forced with \textbackslash\textbackslash
}%

\author{Jingbiao Chen}
\affiliation{ 
State Key Laboratory of Advanced Optical Communication Systems and Networks, Institute of Quantum Electronics, School of Electronics, Peking University, Beijing 100871, China%\\This line break forced with \textbackslash\textbackslash
}%
\date{\today}% It is always \today, today,

\begin{abstract}
We report the development of  laser systems- a “Voigt laser” - using a Voigt anomalous dispersion optical filter as the frequency-selective element, working at the wavelength of 780 nm of $^{87}$Rb-D2 resonance line. Compared with Faraday anomalous dispersion optical filter, the Voigt anomalous dispersion optical filter can generate a stronger and more uniform magnetic field with a compact size of magnet, and obtains a transmission spectrum with narrower linewidth and more stable lineprofile. In this case, the frequency stability of the Voigt laser reaches 5$\times$10$^{-9}$ at the averaging time of 200 s, and the wavelength fluctuation of 8-hours free operation is ± 0.1 pm. Besides, the Voigt laser has greater immunity to diode current than the Faraday laser, with a wavelength fluctuation of ± 0.5 pm in the current range from 73 mA to 150 mA. Finally, the Voigt laser frequency can be controlled by the cell temperature of the Voigt optical filter, which is expected to achieve a frequency detuning of 20 GHz.  Consequently, the Voigt laser, whose frequency could correspond to the atomic transition frequency by tuning the cell temperature, obtains good robustness to the current and temperature fluctuation of laser diode, and could realize a compact optical standard for precise measurement once stabilized by modulation transfer spectroscopy.
\end{abstract}
\maketitle
Diode laser with narrow linewidth plays an essential role in the study of atomic physics, and is used in an extremely broad range of applications from laser cooling to thermal vapor experiments\cite{r1}. In the usual case, the frequency-selective element of a traditional laser could be formed with a spatial grating\cite{r4,r5} or a Fabry–Pérot (FP) etalon\cite{r6}. The frequency of these lasers is sensitive to their diode current and temperature, requiring an extremely high precision temperature and current control. Besides, the laser frequency needs to be finely tuned to the atomic transition line, to achieve frequency stabilization using saturated absorption spectroscopy\cite{r7}, polarization spectroscopy\cite{r8}, modulation transfer spectrum\cite{r9,r10}, etc. In this case, there will be a risk of mode hopping, resulting in the laser frequency far away from the locking point, which is not conducive to long-term continuous work.

Faraday laser comprises a Faraday anomalous dispersion optical filter (FADOF) as a frequency-selective element, could solve the above problem\cite{r18,tao2015diode,tao2016faraday,r3,r2,r21}. FADOF has attracted much attention since its birth\cite{r11}, owing to its narrow bandwidth and high signal-to-noise ratio (SNR)\cite{r12,r13,r14,r15,r16,r17}. Since 2011, FADOF is applied as a frequency-selective element, revealing a stable "Faraday laser" immune to the diode current and temperature\cite{r18}. Importantly, the wavelength of the Faraday laser is limited to the narrow-band transmission window of FADOF, which is around the atomic transition line. Therefore, the Faraday laser could easily achieve the frequency stabilization once it is turned on, and could stably operate for a long term without artificial adjustment\cite{shi2022frequency}. In recent years, the Faraday laser has been formally proposed to different structures and atoms, achieving better frequency stability and anti-interference ability\cite{r27,r2,r19,r20,r21}. Nevertheless, the axial magnetic field in the Faraday laser is generally generated by two magnets aside the cell along the optical axis, which greatly increases the axial length of FADOF and limits the minimum cavity length of Faraday laser\cite{r19}. In another method, the strength and uniformity of the magnetic field generated by the magnet around the cell are relatively small, which limits the further improvement of the laser stability\cite{r2}.  

Unlike the Faraday effect, the Voigt effect rotates the polarization direction of the incident laser by applying a radial magnetic field\cite{r22}. Therefore, the Voigt anomalous dispersion optical filter (VADOF) could construct a stronger and more homogeneous magnetic field with less volume than FADOF, by placing two magnets close together in the radial direction of the cell\cite{r23,r24,r25}. Then, a stable transmission spectrum with a narrower linewidth is obtained. In this case, a compact "Voigt laser" could be constructed, using a VADOF for frequency-selection. Like the Faraday laser, the Voigt laser could realize compact optical frequency standard, using modulation transfer spectroscopy (MTS) for frequency stabilization  \cite{shi2022frequency,miao2022compact}.  

In this work, we propose and prove a Voigt laser, which uses $^{87}$Rb VADOF as frequency-selective element and semiconductor diode laser as gain medium. This Voigt laser operated freely for many times during half a year, and its wavelength fluctuation is ± 0.5 pm with an instability better than $2\times10^{-8}$ for 48 hours. Compared with the Faraday laser, the Voigt laser has a better frequency stability in a smaller volume. Besides,  the wavelength fluctuation of the Voigt laser is ± 0.5 pm in the diode current range from 73 mA to 150 mA, presenting the Voigt laser's immunity towards the parameters of the diode laser . Meanwhile, we investigate the influence of the cell temperature on the Voigt laser, revealing a tunable laser wavelength selected by the cell temperature. Therefore, the  frequency of this Voigt laser can corresponds to the atomic transition frequency by adjusting the cell temperature, and then can be stabilized using MTS technology to realize a compact optical frequency standard. Furthermore, the Voigt laser could use different gain mediums and  atoms to improve its performance, which paves the way for basic scientific research (such as atomic clock, atomic gravimeter, and atomic interferometer, etc.) and industrial electronic equipment (precision optical measurement, spectral inspection).

\begin{figure}[htp]
    \centering
    \includegraphics[width=8cm]{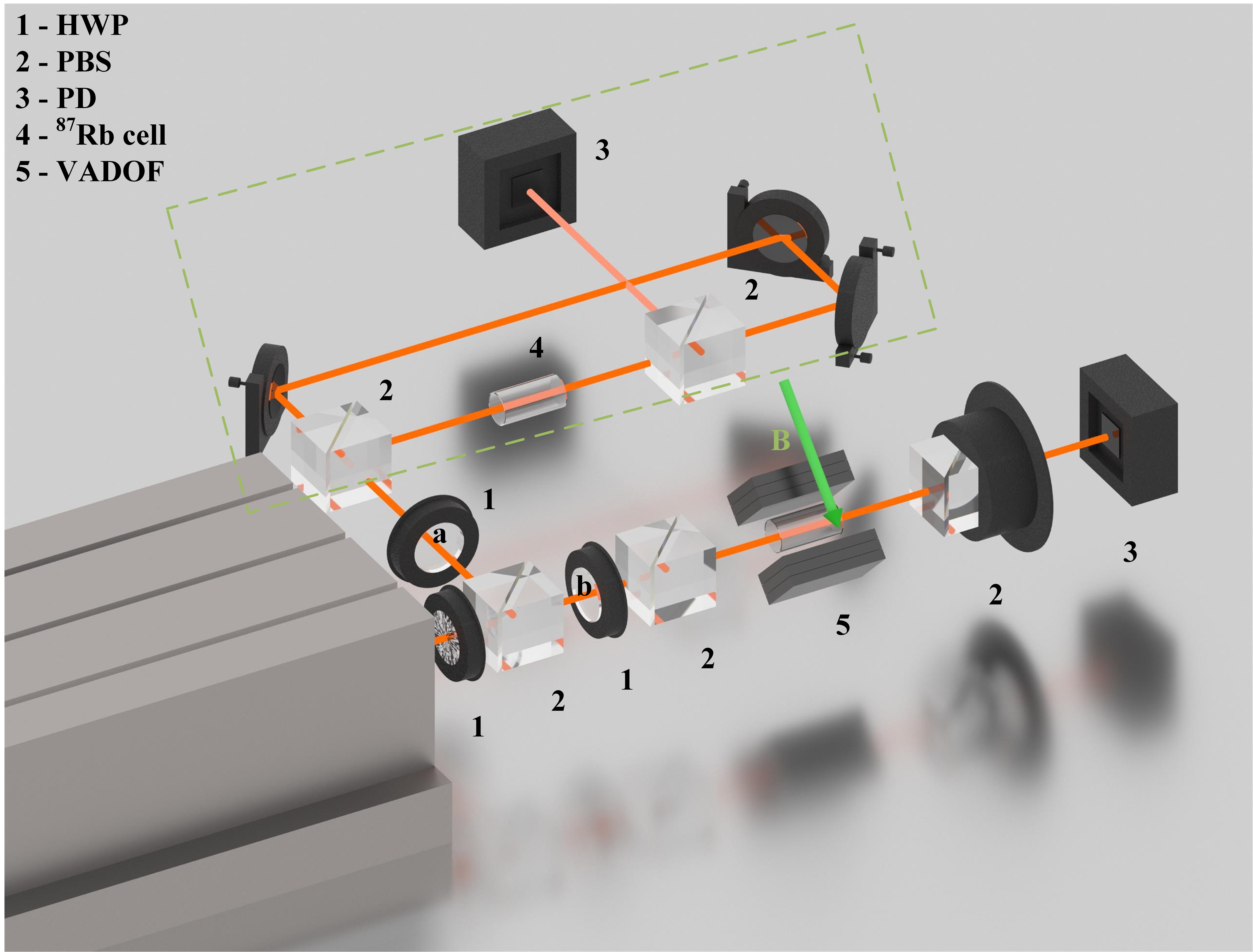}
    \caption{Experimental setup of the VADOF transmission spectrum. The green dash refers to the saturated absorption spectroscopy (SAS). HWP, half-wave plate; PBS, polarization beam splitter; PD, photon detector; The probe laser is split into two laser beams (laser a and laser b)  by a HWP and a PBS. }
    \label{fig1}
\end{figure}

VADOF comprises a $^{87}$Rb atomic vapor cell with length of \textit{L} = 30 mm, a set of permanent magnets, a temperature control system, and two polarized beam splitters (PBSs), as presented in Fig. ~\ref{fig1}. The vacuum cell and magnets are stuck in a desired Teflon thermal insulation structure. The anti-reflection coated cell has a transmission of 95\% for $\lambda$  = 780 nm laser. The temperature of the $^{87}$Rb cell is controlled by an electric heating element that allows to maintain the cell temperature within 0.1℃. The radial magnetic field range from 200 Gs to 3500 Gs, which reach a maximum with two 15 mm-thick NdFeB permanent magnets aside the cell. For a better Voigt effect, there is a 45° angle between the magnetic field direction and the laser polarization direction\cite{r24}. The two PBSs are placed on either side of the cell. The polarization direction of the PBS behind the cell can be rotated, so that the two PBSs could be orthogonal or parallel.

We use a homemade 780 nm interference-filter external cavity diode laser (IF-ECDL) to probe the transmission spectrum of VADOF, of which the laser beam diameter is 2 mm. The probe laser is split into two laser beams (laser a and laser b) by a half wavelength plate (HWP) and a PBS. We probe the saturated absorption spectroscopy (SAS) of $^{87}$Rb as a frequency reference using laser a, as the green dash in Fig. ~\ref{fig1} shows. Laser b goes through a HWP to control the laser intensity before probing the transmission spectrum of VADOF. The above spectrums are simultaneously presented in an oscilloscope, which connects with two photodetectors (PD, Thorlabs PDA8A2).

\begin{figure}[htp]
    \centering
    \includegraphics[width=8cm]{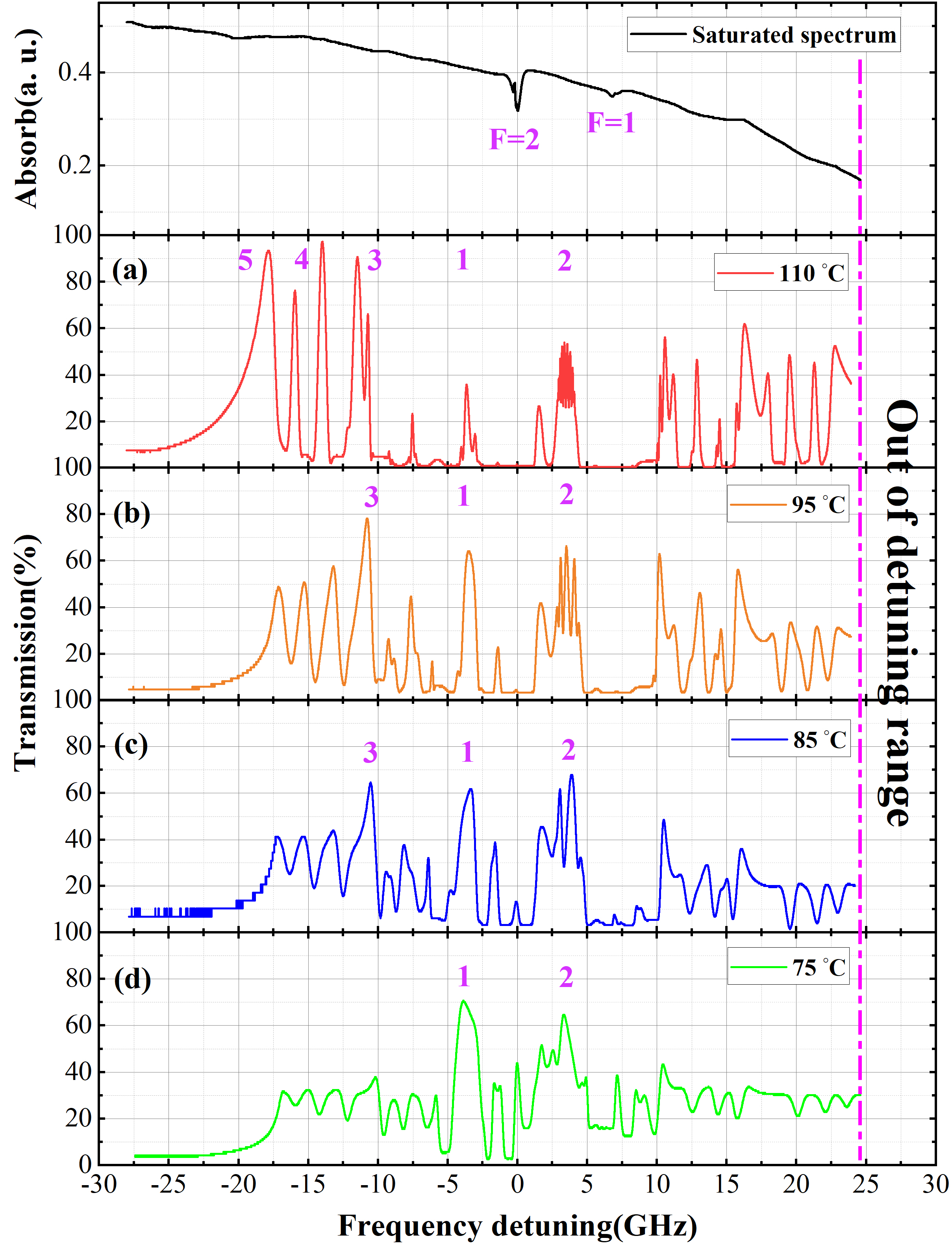}
    \caption{The saturated absorption spectroscopy of $^{87}$Rb-D2-line and the transmission spectrum of VADOF in different temperature.The diagram at the top shows  \textit{F} = 1 and \textit{F} = 2 transition of $^{87}$Rb-D2-line in SAS.  a, b ,c , d present the transmission spectrum at the cell temperature of 110℃, 95℃, 85℃, and 75℃, respectively. Peak 1, 2, 3, 4, and 5 are typical transmission peaks in the transmission spectrum. The probe laser frequency detuning ranges from -30 GHz to 25 GHz.}
    \label{fig2}
\end{figure}  

Before experimentally probing the spectrum of VADOF, we analyze its major influencing factors using a simulation tool\cite{r26}. The results show that we can reveal a spectrum with high transmission and narrow linewidth, when radial magnetic field \textit{B} = 3500 Gs. Then, we investigate the transmission spectrum of VADOF in different temperatures at \textit{B} = 3500 Gs, as shown in Fig. ~\ref{fig2}. The diagram at the top shows SAS of $^{87}$Rb-D2-line. The frequency interval between \textit{F} = 1 and \textit{F} = 2 transition is 6.834 GHz, and we set the frequency at the peak of \textit{F} = 2 transition as the detuning origin. 

The probe laser frequency tuning ranges from -30 GHz to 25 GHz, where we can almost observe the whole transmission spectrum. Limited by the tuning range, The spectrum above 25 GHz is missed. However, the transmission spectrum above 25 GHz is suggested to monotonically decrease to 0 (the same trend as the left side of the transmission spectrum), according to the theoretical prediction. Compared with conventional FADOF, VADOF requires higher temperatures to reveal high transmission. Therefore, the Voigt laser is hard to produce once the cell temperature is lower than 60℃. When the cell temperature is in the range of 60 - 80℃, the peak 1 in  the transmission spectrum obtains the highest transmission, seeing Fig. ~\ref{fig2}d. In this case, the transmission peak 1, 2, and 3 will rise with the cell temperature. Peaks 2 and 3 become the highest transmission peak at 85℃ and 95℃, respectively, as presented in Fig. ~\ref{fig2}c and Fig. ~\ref{fig2}b. VADOF reveals a ‘wing’ spectral profile and peak 4 stands out as the highest, as the temperature rises to 110℃ in Fig. ~\ref{fig2}a. Then, the "wing" peak 5 will become the highest and the spectral profile won’t change with the rising temperature. However, the linewidth of peak 5 is relatively wide and its center frequency is sensitive to the cell temperature, which is not suitable for Voigt laser. Therefore, the highest transmission peak gradually jumps from peak 1 to peak 4 when the temperature increases from 60°C to 110°C. Based on this, we predict that the output laser wavelength of the Voigt laser also undergoes corresponding jumps when the temperature increases. 

  \begin{figure}[htp]
    \centering
    \includegraphics[width=8cm]{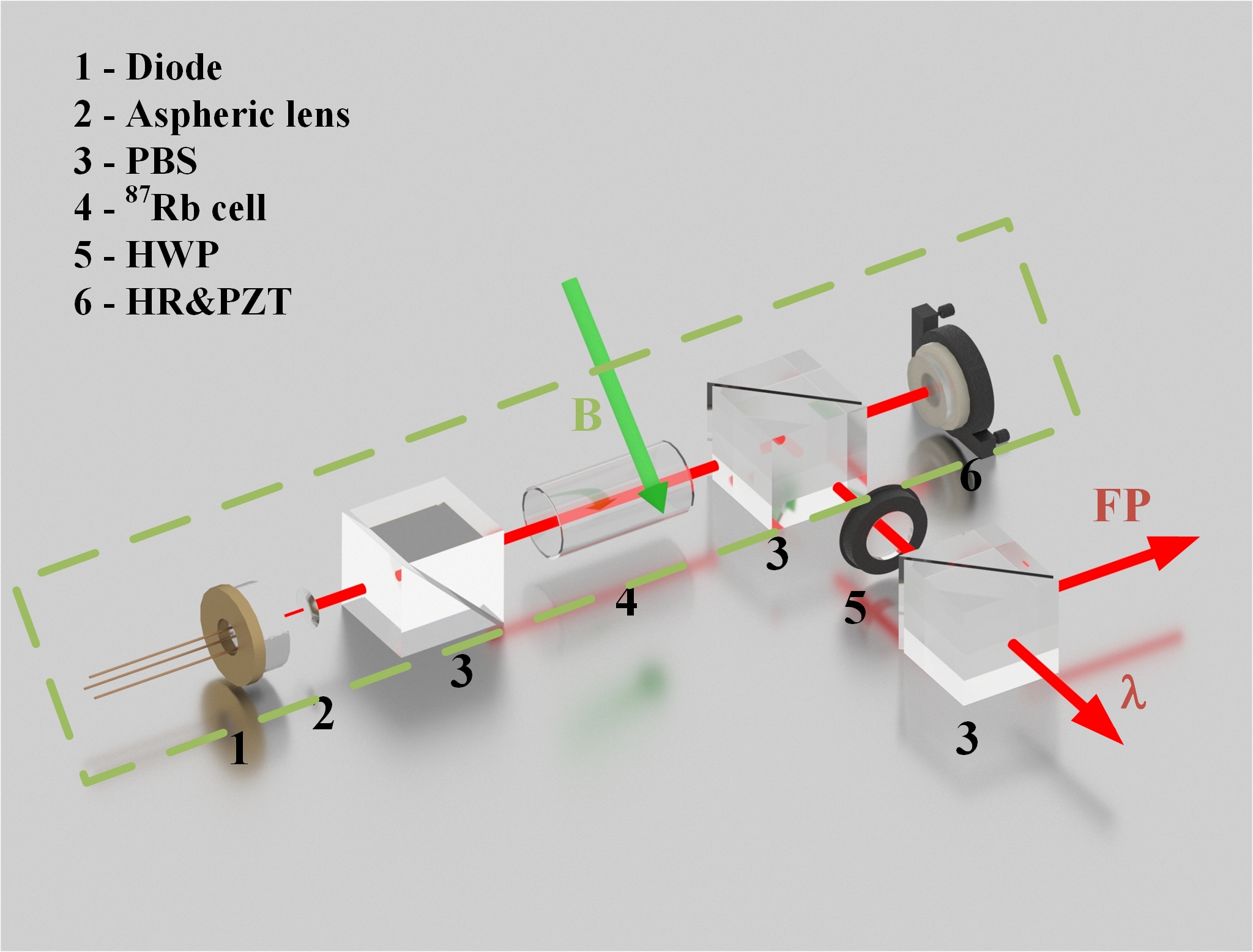}
    \caption{The experiment setup of the Voigt laser. HWP, half-wave plate; PBS, polarization beam splitter; HR, high-reflection mirror; PZT, piezoelectric ceramic tube; FP, Scanning Fabry–Pérot Interferometers (Thorlabs, SA210-5B); $\lambda$, optical wavelength meter (Bristol 671A)}
    \label{fig3}
\end{figure} 

The experimental setup of the Voigt laser is schematically shown in the green dash of Fig. ~\ref{fig3}. The Voigt laser with the length of optical cavity of 20 cm is composed of a $^{87}$Rb VADOF,  an anti-reflection coated laser diode, aspheric lens, a high-reflection mirror (HR) and a piezoelectric ceramic tube (PZT). After collimated by an aspheric lens, the beam diameter of the ARLD’s emitted light is enlarged to 2 mm. Then, the collimated light is filtered by the Voigt optical filter and fed back into the ARLD by HR. The reflectivity of HR is 99.9\% at the wavelength of 780 nm, and is fixed on the PZT for finely detuning the cavity length. The laser beam outputs from the second PBS. A HWP and a PBS are comprised to split the output laser, for scanning FP spectrum and sensing laser wavelength, respectively. The temperature of $^{87}$Rb cell could range from 20℃ to 200℃ and the radial magnetic field is between 200 Gs and 3500 Gs.

\begin{figure}[htp]
    \centering
    \includegraphics[width=8cm]{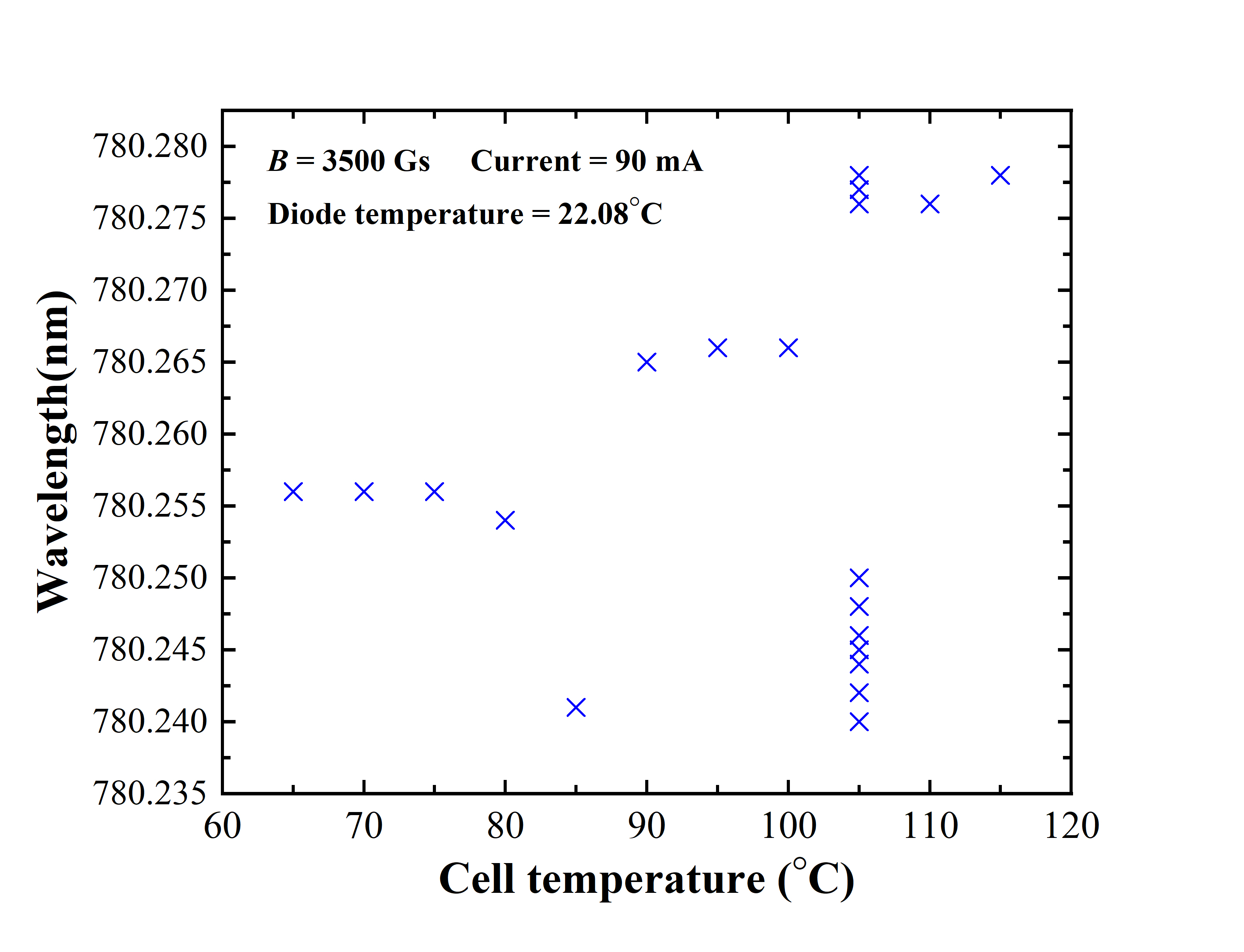}
    \caption{The wavelength of the Voigt laser at different cell temperature.}
    \label{fig4}
\end{figure}

We firstly investigate the cell temperature’s effect on the Voigt laser, presented in Fig. ~\ref{fig4}. The radial magnetic field is selected as 3500 Gs for a high transmission peak. The LD current and temperature are 90 mA and 22.08℃, respectively. The output laser wavelength jumps with the increasing cell temperature, as we predicted in the discussion of Fig. ~\ref{fig2}. Once the temperature is stabilized at an exact value, laser with a stable wavelength will be emitted. When the temperature varies between 60 - 80℃, the output laser wavelength is maintained at 780.256 nm (near 60 – 75℃) or 780.254 nm (near 80℃), corresponding to the two longitudinal modes of peak 1 in Fig. ~\ref{fig2}. Continuing to increase the cell temperature, the laser wavelength will be detected as 780.241 nm (near 85℃) and then jump to 780.265 nm (near 90-100℃), referring to peak 2 and peak 3 in Fig. ~\ref{fig2}, respectively. When the temperature is near 105℃, the transmission of peaks in Fig. ~\ref{fig2} are close to each other and the output laser wavelength fluctuates widely. In this case, the laser wavelength will level out at 780.276 nm, corresponding to peak 4 in Fig. \ref{fig2}, if the cell temperature rises to 110℃. Then, the output laser will be red-shifted and the laser wavelength will rise with the increasing cell temperature. Subsequently, the Voigt laser with an EOM could achieve a wide frequency tuning range of 20 GHz. 

\begin{figure}[htp]
    \centering
    \includegraphics[width=8cm]{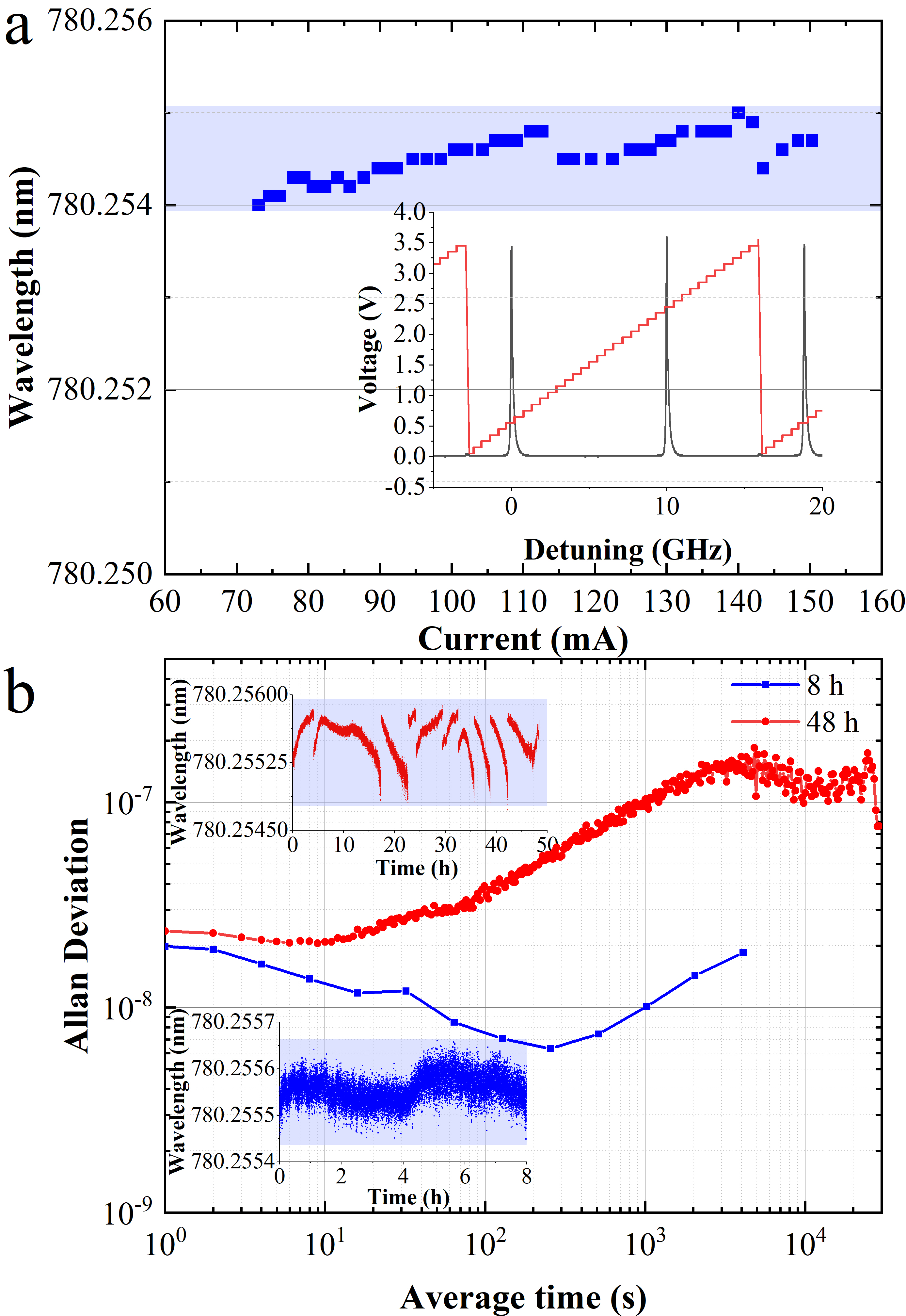}
    \caption{The characteristics of Voigt laser. \textbf{a} laser wavelength versus LD current and the result of Scanning Fabry–Pérot Interferometers. \textbf{b} Allan deviation of laser frequency for 8-hour (blue solid) and 48-hour (red solid) operation. The raw data is shown in the inset ( the blue dotted curve for 8-hour operation and the red dotted curve for 48-hour operation), indicating the wavelength variation during  operation.}
    \label{fig5}
\end{figure}

Based on the above investigation, we further study the characteristics of the Voigt laser at different LD currents, seeing Fig. ~\ref{fig5}. The cell temperature is set to 75℃. The blue square in Fig. ~\ref{fig5}a shows the laser wavelength at different LD currents ranging from 73 mA to 150 mA. The wavelength of the output laser is near 780.255 nm and the wavelength fluctuation is about ± 0.5 pm. Furthermore, the Voigt laser maintains a single longitudinal mode when the LD current ranges from 73 mA to 150 mA, as shown in the insert of Fig. ~\ref{fig5}a. For comparison, the LD current range of Faraday laser without mode-hopping is 74 – 130 mA and the wavelength fluctuation is ± 2 pm\cite{r2}. Therefore, the Voigt laser has stronger immunity to current drift. In addition, our preliminary experimental results show that the Voigt laser is also immune to the diode temperature.

With the LD current set as 150 mA, we investigate the frequency instability for a long-term free operation. We record the wavelength of the output laser using a high-precision optical wavelength meter (Bristol 671A), at a sampling rate of 1 ~Hz. As shown in the insert of Fig. ~\ref{fig5}b, the wavelength fluctuation  is ± 0.1 pm and ± 0.5 pm for 8-hour and 48-hour operation, respectively. However, the wavelength fluctuation is ± 2 pm for 48-hour operation in Faraday laser\cite{r2}. We then analyze this data set using the standard approach of an Allan deviation, obtaining the long-term instability of the Voigt laser, as presented in Fig. ~\ref{fig5}b. For a 8-hour free-running, the frequency instability of a  Voigt laser can reach $5\times10^{-9}$ as the average time is up to 200 s, which is on the order of the Faraday laser using cavity locking technology\cite{r3}. When the operation time raises to  48 hours, the frequency instability can still reach $2\times10^{-8}$, even if  affected by  the external vibration and the ambient temperature variation for over 10 degrees.  Consequently, the Voigt laser, with excellent frequency stability and immunity to the diode laser current and temperature, could contribute to the development of the compact optical frequency standard for the application of atomic clock, atomic gravimeter and outdoor precision measurement experiment.

In Conclusion, this letter put forward a novel implementation scheme for diode laser, named as a "Voigt laser", using the Voigt effect for frequency selection. The Voigt laser obtains a frequency instability of 5$\times$10$^{-9}$ and  a strong immunity to the temperature and current of the laser diode. Moreover, the cell temperature characteristics of the Voigt laser in a large magnetic field are explored. Results show that the Voigt laser wavelength could be selected by adjusting the cell temperature and the frequency variation range is 20 GHz. Therefore, the compact Voigt laser,  corresponding to the atomic transition line by adjusting the cell temperature, can serve for the atomic clocks, atomic gravimeters, and atomic gyroscopes, owing to to its high frequency stability and stable long-term operation. Importantly, the Voigt laser using different gain mediums and atoms,  could be applied to active optical clock\cite{zhuang2014active} and cold atom optical clock in the following study, promoting the development of optical frequency standards.  

 \section*{ACKNOWLEDGMENTS}
The work was was funded by the National Natural Science Foundation of China (NSFC) (91436210), China Postdoctoral Science Foundation (BX2021020), Wenzhou Major Science and Technology Innovation Key Project (ZG2020046), and Wenzhou Key Scientific and Technological Innovation R\&D Project (2019ZG0029).

\section*{Data Availability Statement}

The data that support the findings of
this study are available from the
corresponding author upon reasonable
request.

\bibliography{Ref}

\end{document}